\documentclass[12pt,titlepage,a4paper]{article}

\usepackage{amssymb}

\advance\textwidth1.5cm
\advance\textheight1cm
\advance\oddsidemargin-1.2cm
\advance\topmargin-1cm

\newcommand{\UH}{\widehat{\mathfrak U}}

\newcommand{\HB}{\widehat{B}}
\newcommand{\HJ}{\widehat{J}}
\newcommand{\HS}{\widehat{S}}
\newcommand{\HT}{\widehat{T}}

\newcommand{\HJQ}{\HJ \!\!\!   \backslash}
\newcommand{\HBQ}{\HB \!\!\!   \backslash}
\newcommand{\HSQ}{\HS \!\!\!   \backslash}
\newcommand{\HTQ}{\HT \!\!\!\! \backslash}

\newcommand{\gah}{\widehat{\mathfrak a}}

\newcommand{\Zet}{\mathbb{Z}}

\newcommand{\ubr}{\linebreak[0]}
\newcommand{\ubrii}{\linebreak[0]}
\newcommand{\br}{\linebreak[4]}

\begin{document}

\begin{titlepage}

\begin{flushright} { THEP 98/16\\University of Freiburg\\July 1998}
\end{flushright}
\vspace{8mm}

\begin{center}
{\large \bf The Nambu--Goto Theory of Closed Bosonic Strings\\[2mm] 
            Moving in $1+3$--Dimensional Minkowski Space: \\[2mm]
            The Construction of the Quantum Algebra of Observables\\[2mm] 
            Up to Degree Five}\\[1cm] { G. Handrich and C. 
Nowak}\\[3mm] {Fakult\"at f\"ur Physik der Universit\"at Freiburg, 
Hermann--Herder--Str.\ 3, \\ D--79104 Freiburg, Germany}\\[2.5cm] 
{Addendum to hep--th/9805057:}\\[4mm] 

{\large The Nambu--Goto Theory of Closed Bosonic Strings\\ 
           Moving in $1+3$--Dimensional Minkowski Space: \\ The 
Quantum Algebra of Observables\\[3mm]}{by K. Pohlmeyer}\\[2cm] {\bf 
Abstract}\\[7mm] 

\end{center}

\noindent
The quantum algebra of observables postulated in 
hep--th/9805057 is constructed up to degree {\em five}. All independent 
relations of degree four are given; they involve three as yet undetermined 
parameters. Definitions and symbols are used as introduced in the 
above-mentioned article.

\end{titlepage}

In the course of constructing the quantum algebra of observables of  
closed bosonic strings moving in $1+3$ dimensions the cycle of degree four 
of the deformation routine has been carried out. This means that all the 
relations of degree four have been properly taken into account, and that the 
resulting algebra consistently satisfies all correspondence postulates in 
degrees five and lower. Due to the immense computational effort involved in 
this cycle, all the necessary operations were performed with the help of 
Mathematica routines specially designed to take advantage of the $so(3)$ 
representation space structure. 

In order to find all the implications in degree five of the truly independent
relations of degrees four and lower, in addition to the $\UH$-relations 
of degree four obtained in the preceding cycle, four further relations
have to be taken into account: One relation defining the maximal abelian 
algebra $\gah$: $[\HBQ_0^{(3)},\HBQ_0^{(1)}]_0=0$, and three defining
the action of the element $\HBQ_0^{(3)}$ on the subalgebra $\UH$. The
latter are obtained from the corresponding relations of the classical
algebra of observables by means of the steps i) -- iv). Before having
established consistency in degree five, these relations involve 60
new parameters, 9 of order $\hbar^2$, 47 and 4 of order $\hbar$ and
$\hbar^3$ respectively. Moreover, there remains one further parameter in
the $\UH$-relations, which has not been determined in the previous
cycles, namely f. 

As it turns out --- remember that the dimension of each subspace 
homogeneous with respect to degree, spin and parity is explicitly known 
--- there is one further truly independent relation in the spin-parity 
channel $0^+$. All other $\UH$-relations of degree five follow from the 
$\UH$-relations of degree four without the use of $\HBQ_0^{(3)}$ and 
$\HBQ_0^{(1)}$. Therefore, the postulate that there are no relations
without a classical counterpart entails that all relations induced to
degree five from the relations of degree four which involve
$\HBQ_0^{(3)}$ are linearly dependent on the already obtained relations.
Linear dependence disregarding terms of degree lower than five is
ensured by construction. By the said postulate, the coefficients of lower
degree terms must vanish, thus giving restrictions on the 61 parameters. 

To be explicit, the relation $[\HBQ_0^{(3)},\HBQ_0^{(1)}]_0=0$ is promoted 
to degree five, spin-parity $1^-$, $2^-$ and $2^+$ with the help of $\HSQ_1$,
$\HSQ_2$ and $\HTQ_2$ respectively, and the truly independent relations of
degree two are promoted to degree five with the help of $\HBQ_0^{(3)}$,
producing relations of spin-parity $J^P=4^-$, $3^+$ (two), $3^-$, $2^-$ and
$1^+$. All in all, a total number of 569 coefficients of terms of lower
degree than five are required to vanish, each coefficient being, up to a 
common real or imaginary square root, a polynomial in the parameters with
rational coefficients. 

At first considering the linear homogeneous equations, it turns out that,
as in the cycle before, all parameters of first order in Planck's constant
must be equal to zero. Using this result in the other equations, the ones 
obtained by the vanishing of the terms of degree two involve parameters of
third order in $\hbar$ only. They are homogeneous and imply the 
vanishing of all these parameters. Amongst the now linear inhomogeneous 
equations involving the parameters of second order in $\hbar$, seven turn 
out to be independent. The equations implied by the vanishing of the 
coefficients in front of terms of degree one are inhomogeneous; they involve
products of parameters of order $\hbar^2$. It turns out that these
equations are satisfied trivially, as soon as one reduces the system with
the help of the equations obtained before. Hence, requiring correspondence
and consistency up to degree five leaves three parameters undetermined:
besides f two suitably chosen parameters $g_1$ and $g_2$. 

The vanishing of all parameters of odd degree in Planck's 
constant supports the conjecture that a $\Zet_2$-grading will survive the  
construction process of the quantum algebra of observables. As in the previous 
cycles, all equations giving independent restrictions on the parameters are 
linear. Therefore, we still expect the parameters, initially defined to 
represent real numbers, to be rational-valued indeed.   

Here are the relations defining the action of $\HBQ_0^{(3)}$ on the 
subalgebra $\UH$ with  the remaining parameters in them: 

\begin{sloppy}
\begin{flushleft}

\hangindent=5mm
\hangafter=1
$J^P=2^+$:\\[3mm]
$
[\HBQ_0^{(3)},\HTQ_{2}]_{2} = \ubr- i\,20\ubrii\,{\sqrt{{\frac{2}{3}}}}\ubrii\,
      [[[\HSQ_{2},\HSQ_{1}]_{1},\HTQ_{2}]_{1},\HTQ_{2}]_{2}  \ubr+ 
   {\frac{35}{6}}\ubrii\,{\sqrt{{\frac{5}{3}}}}\ubrii\,
    [[[\HTQ_{2},\HSQ_{1}]_{2},\HSQ_{1}]_{1},\HTQ_{2}]_{2} \ubr+ 
   {\frac{6247}{162}}\ubrii\,{\sqrt{{\frac{1}{35}}}}\cdot\ubrii\,
    [[[\HTQ_{2},\HSQ_{1}]_{1},\HSQ_{1}]_{2},\HTQ_{2}]_{2} \ubr+ 
   {\frac{2749}{81}}\ubrii\,{\sqrt{{\frac{1}{5}}}}\ubrii\,
    [[[\HTQ_{2},\HSQ_{1}]_{1},\HSQ_{1}]_{1},\HTQ_{2}]_{2} \ubr- 
   {\frac{239}{324}}\ubrii\,{\sqrt{5}}\ubrii\,[[[\HTQ_{2},\HSQ_{1}]_{1},\HSQ_{1}]_{0},\HTQ_{2}]_{2} 
   \ubr- {\frac{983}{5}}\ubrii\,{\sqrt{{\frac{1}{210}}}}\ubrii\, 
    [[\HTQ_{2},\HSQ_{1}]_{3},[\HTQ_{2},\HSQ_{1}]_{2}]_{2} \ubr- 
   {\frac{62407}{810}}\ubrii\,{\sqrt{{\frac{1}{21}}}}\ubrii\,
    [[\HTQ_{2},\HSQ_{1}]_{3},[\HTQ_{2},\HSQ_{1}]_{1}]_{2} \br- 
   {\frac{43688}{405}}\ubrii\,{\sqrt{{\frac{1}{15}}}}\ubrii\,
    [[\HTQ_{2},\HSQ_{1}]_{2},[\HTQ_{2},\HSQ_{1}]_{1}]_{2} \ubr+ 
   i\,{\frac{152291}{540}}\ubrii\,{\sqrt{{\frac{1}{105}}}}\ubrii\,
    [[[\HSQ_{2},\HSQ_{1}]_{3},\HSQ_{1}]_{3},\HSQ_{1}]_{2} \br- 
   i\,{\frac{307087}{540}}\ubrii\,{\sqrt{{\frac{1}{210}}}}\ubrii\,
    [[[\HSQ_{2},\HSQ_{1}]_{2},\HSQ_{1}]_{3},\HSQ_{1}]_{2} \ubr+ 
   i\,{\frac{335}{18}}\ubrii\,{\sqrt{{\frac{1}{6}}}}\ubrii\,
    [[[\HSQ_{2},\HSQ_{1}]_{2},\HSQ_{1}]_{2},\HSQ_{1}]_{2} \br+ 
   i\,{\frac{38849}{540}}\ubrii\,{\sqrt{{\frac{1}{10}}}}\ubrii\,
    [[[\HSQ_{2},\HSQ_{1}]_{2},\HSQ_{1}]_{1},\HSQ_{1}]_{2} \ubr- 
   i\,{\frac{2861}{108}}\ubrii\,{\sqrt{{\frac{5}{2}}}}\ubrii\,
    [[[\HSQ_{2},\HSQ_{1}]_{1},\HSQ_{1}]_{2},\HSQ_{1}]_{2} \br+ 
   i\,{\frac{421}{30}}\ubrii\,{\sqrt{{\frac{1}{30}}}}\ubrii\,
    [[[\HSQ_{2},\HSQ_{1}]_{1},\HSQ_{1}]_{1},\HSQ_{1}]_{2} \ubr+
   i\,{\frac{187973}{324}}\ubrii\,{\sqrt{{\frac{1}{21}}}}\ubrii\,
    \{\HJQ_{1},[[\HSQ_{2},\HSQ_{1}]_{1},\HTQ_{2}]_{3}\}_{2} \br+ 
   {\frac{1074869}{270}}\ubrii\,{\sqrt{{\frac{1}{210}}}}\ubrii\,
    \{\HJQ_{1},[[\HTQ_{2},\HSQ_{1}]_{2},\HSQ_{1}]_{3}\}_{2} \ubr- 
   {\frac{669}{10}}\ubrii\,{\sqrt{{\frac{1}{10}}}}\ubrii\,
    \{\HJQ_{1},[[\HTQ_{2},\HTQ_{2}]_{1},\HTQ_{2}]_{2}\}_{2} \br+ 
   {\frac{7558769}{8100}}\ubrii\,{\sqrt{{\frac{1}{210}}}}\ubrii\,
    \{\HJQ_{1},[[\HTQ_{2},\HSQ_{1}]_{3},\HSQ_{1}]_{2}\}_{2} \ubr- 
   {\frac{6918529}{48600}}\ubrii\,{\sqrt{{\frac{1}{6}}}}\ubrii\,
    \{\HJQ_{1},[[\HTQ_{2},\HSQ_{1}]_{2},\HSQ_{1}]_{2}\}_{2} \br+ 
   {\frac{50923}{5400}}\ubrii\,{\sqrt{{\frac{1}{10}}}}\ubrii\,
    \{\HJQ_{1},[[\HTQ_{2},\HSQ_{1}]_{1},\HSQ_{1}]_{2}\}_{2} \ubr- 
   i\,{\frac{42617}{1800}}\ubrii\,{\sqrt{{\frac{1}{21}}}}\ubrii\,
    \{\HJQ_{1},[[\HSQ_{2},\HSQ_{1}]_{2},\HTQ_{2}]_{2}\}_{2} \br+ 
   i\,{\frac{399187}{1080}}\ubrii\,{\sqrt{{\frac{1}{15}}}}\ubrii\,
    \{\HJQ_{1},[[\HSQ_{2},\HSQ_{1}]_{1},\HTQ_{2}]_{2}\}_{2} \ubr+ 
   {\frac{675377}{1080}}\ubrii\,{\sqrt{{\frac{1}{10}}}}\ubrii\,
    \{\HJQ_{1},[[\HTQ_{2},\HSQ_{1}]_{2},\HSQ_{1}]_{1}\}_{2} \br- 
   {\frac{753551}{3240}}\ubrii\,{\sqrt{{\frac{1}{30}}}}\ubrii\,
    \{\HJQ_{1},[[\HTQ_{2},\HSQ_{1}]_{1},\HSQ_{1}]_{1}\}_{2} \ubr- 
   i\,{\frac{972803}{1080}}\ubrii\,{\sqrt{{\frac{1}{15}}}}\ubrii\,
    \{\HJQ_{1},[[\HSQ_{2},\HSQ_{1}]_{2},\HTQ_{2}]_{1}\}_{2} \br+ 
   i\,{\frac{311647}{3240}}\ubrii\,\{\HJQ_{1},[[\HSQ_{2},\HSQ_{1}]_{1},\HTQ_{2}]_{1}\}_{2} 
   \ubr- i\,{\frac{60572}{27}}\ubrii\,{\sqrt{{\frac{1}{35}}}}\ubrii\, 
    \{[\HSQ_{2},\HSQ_{1}]_{3},\HTQ_{2}\}_{2} \ubr- 
   i\,{\frac{87811}{270}}\ubrii\,{\sqrt{{\frac{1}{14}}}} \cdot \ubrii\,
    \{[\HSQ_{2},\HSQ_{1}]_{2},\HTQ_{2}\}_{2} \ubr- 
   {\frac{59}{2}}\ubrii\,{\sqrt{{\frac{5}{3}}}}\ubrii\,\{[\HTQ_{2},\HTQ_{2}]_{1},\HTQ_{2}\}_{2} 
   \ubr+ i\,{\frac{613}{27}}\ubrii\,{\sqrt{{\frac{1}{10}}}}\ubrii\, 
    \{[\HSQ_{2},\HSQ_{1}]_{1},\HTQ_{2}\}_{2} \ubr+ 
   {\frac{2218}{27}}\ubrii\,{\sqrt{{\frac{1}{3}}}} \cdot \ubrii\,
    \{[\HSQ_{1},\HSQ_{1}]_{1},\HTQ_{2}\}_{2} \ubr+ 
   i\,{\frac{278041}{270}}\ubrii\,{\sqrt{{\frac{1}{35}}}}\ubrii\,
    \{[\HTQ_{2},\HSQ_{1}]_{3},\HSQ_{2}\}_{2} \ubr+ 
   i\,{\frac{79091}{135}}\ubrii\,{\sqrt{{\frac{1}{14}}}}\ubrii\,
    \{[\HTQ_{2},\HSQ_{1}]_{2},\HSQ_{2}\}_{2} \ubr- 
   {\frac{2021}{9}} \cdot \ubrii\,{\sqrt{{\frac{5}{3}}}}\ubrii\,
    \{[\HTQ_{2},\HSQ_{2}]_{1},\HSQ_{2}\}_{2} \ubr- 
   i\,{\frac{3748}{15}}\ubrii\,{\sqrt{{\frac{2}{5}}}}\ubrii\,
    \{[\HTQ_{2},\HSQ_{1}]_{1},\HSQ_{2}\}_{2} \ubr- 
   {\frac{359}{10}}\ubrii\,{\sqrt{{\frac{1}{5}}}}\ubrii\,
    \{[\HTQ_{2},\HSQ_{2}]_{0},\HSQ_{2}\}_{2} \ubr+ 
   {\frac{1838749}{1620}} \cdot \ubrii\,{\sqrt{{\frac{1}{35}}}}\ubrii\,
    \{[\HTQ_{2},\HSQ_{1}]_{3},\HSQ_{1}\}_{2} \ubr+ 
   {\frac{177263}{2430}}\ubrii\,\{[\HTQ_{2},\HSQ_{1}]_{2},\HSQ_{1}\}_{2} \ubr- 
   i\,{\frac{593}{27}}\ubrii\,{\sqrt{{\frac{5}{2}}}}\ubrii\,
    \{[\HTQ_{2},\HSQ_{2}]_{1},\HSQ_{1}\}_{2} \ubr- 
   {\frac{74}{45}}\ubrii\,{\sqrt{{\frac{1}{15}}}} \cdot \ubrii\,
    \{[\HTQ_{2},\HSQ_{1}]_{1},\HSQ_{1}\}_{2} \ubr- 
   i\,{\frac{597663217}{54675}}\ubrii\,{\sqrt{{\frac{1}{35}}}}\ubrii\,
    \{(\HJQ_{1}^{2})_{2},[\HSQ_{2},\HSQ_{1}]_{3}\}_{2} \ubr+ 
   i\,{\frac{185599217}{109350}}\ubrii\,{\sqrt{{\frac{1}{14}}}} \cdot \ubrii\,
    \{(\HJQ_{1}^{2})_{2},[\HSQ_{2},\HSQ_{1}]_{2}\}_{2} \ubr+ 
   {\frac{192757}{270}}\ubrii\,{\sqrt{{\frac{1}{15}}}}\ubrii\,
    \{(\HJQ_{1}^{2})_{2},[\HTQ_{2},\HTQ_{2}]_{1}\}_{2} \ubr- 
   i\,{\frac{14061367}{12150}}\ubrii\,{\sqrt{{\frac{1}{10}}}} \cdot \ubrii\,
    \{(\HJQ_{1}^{2})_{2},[\HSQ_{2},\HSQ_{1}]_{1}\}_{2} \ubr- 
   {\frac{29309}{135}}\ubrii\,{\sqrt{{\frac{1}{3}}}}\ubrii\,
    \{(\HJQ_{1}^{2})_{2},[\HSQ_{1},\HSQ_{1}]_{1}\}_{2} \ubr- 
   i\,{\frac{43837343}{54675}}\ubrii\,{\sqrt{{\frac{1}{2}}}}\ubrii\,
    \{(\HJQ_{1}^{2})_{0},[\HSQ_{2},\HSQ_{1}]_{2}\}_{2} \ubr+ 
   i\,{\frac{9106621}{3645}}\ubrii\,{\sqrt{{\frac{1}{35}}}}\ubrii\,
    \{\HJQ_{1},\{\HSQ_{2},\HSQ_{1}\}_{3}\}_{2} \ubr- 
   {\frac{793673}{675}}\ubrii\,{\sqrt{{\frac{2}{7}}}}\ubrii\,
    \{\HJQ_{1},(\HTQ_{2}^{2})_{2}\}_{2} \ubr- 
   {\frac{2242283}{675}}\ubrii\,{\sqrt{{\frac{1}{14}}}}\ubrii\,
    \{\HJQ_{1},(\HSQ_{2}^{2})_{2}\}_{2} \ubr+ 
   i\,{\frac{9492257}{14580}}\ubrii\,\{\HJQ_{1},\{\HSQ_{2},\HSQ_{1}\}_{2}\}_{2} \ubr- 
   {\frac{155092}{135}}\ubrii\,{\sqrt{{\frac{2}{3}}}}\ubrii\,
    \{\HJQ_{1},(\HSQ_{1}^{2})_{2}\}_{2} \ubr- 
   i\,{\frac{3616067}{1620}}\ubrii\,{\sqrt{{\frac{1}{15}}}} \cdot \ubrii\,
    \{\HJQ_{1},\{\HSQ_{2},\HSQ_{1}\}_{1}\}_{2} \ubr- 
   {\frac{366608384}{6075}}\ubrii\,{\sqrt{{\frac{1}{35}}}}\ubrii\,
    \{(\HJQ_{1}^{3})_{3},\HTQ_{2}\}_{2} \ubr+ 
   {\frac{100560616}{30375}}\ubrii\,{\sqrt{2}}\ubrii\,
    \{\{(\HJQ_{1}^{2})_{0},\HJQ_{1}\}_{1},\HTQ_{2}\}_{2} \ubr+ i\,\left( 
   {\frac{229}{40}}\,f\ + g_1 + 
   {\frac{10474483}{109350}}\right)\ubrii\,{\sqrt{6}}\ubrii\,[\HSQ_{2},\HSQ_{1}]_{2} 
   \ubr+ \left( {\frac{10631}{600}}\,f + g_2 - 
   {\frac{143384201}{10125}}\right)\ubrii\,{\sqrt{6}}\ubrii\,\{\HJQ_{1},\HTQ_{2}\}_{2}\, , 
$\\[3mm]

$J^P=2^-$:\\*[3mm]


$
[\HBQ_0^{(3)},\HSQ_{2}]_{2} = i\,11\ubrii\,{\sqrt{{\frac{2}{15}}}}\ubrii\,
    [[[\HTQ_{2},\HSQ_{1}]_{2},\HTQ_{2}]_{0},\HTQ_{2}]_{2} \ubr- 
   i\,{\frac{181}{9}}\ubrii\,{\sqrt{{\frac{10}{21}}}}\ubrii\,
    [[[\HTQ_{2},\HSQ_{1}]_{1},\HTQ_{2}]_{2},\HTQ_{2}]_{2} \ubr- 
   i\,{\frac{145}{9}}\ubrii\,{\sqrt{{\frac{2}{3}}}} \cdot \ubrii\,
    [[[\HTQ_{2},\HSQ_{1}]_{1},\HTQ_{2}]_{1},\HTQ_{2}]_{2} \ubr- 
   {\frac{8}{9}}\ubrii\,{\sqrt{5}}\ubrii\,[[[\HSQ_{2},\HSQ_{1}]_{1},\HSQ_{1}]_{0},\HTQ_{2}]_{2} 
   \ubr- {\frac{20684}{189}}\ubrii\,{\sqrt{{\frac{2}{105}}}}\ubrii\, 
    [[\HTQ_{2},\HSQ_{1}]_{2},[\HSQ_{2},\HSQ_{1}]_{3}]_{2} \ubr- 
   {\frac{118}{63}}\ubrii\,{\sqrt{{\frac{1}{21}}}}\ubrii\,
    [[\HTQ_{2},\HSQ_{1}]_{2},[\HSQ_{2},\HSQ_{1}]_{2}]_{2} \ubr+ 
   {\frac{2362}{189}}\ubrii\,{\sqrt{{\frac{1}{15}}}}\ubrii\,
    [[\HTQ_{2},\HSQ_{1}]_{2},[\HSQ_{2},\HSQ_{1}]_{1}]_{2} \ubr+ 
   {\frac{42857}{378}}\ubrii\,{\sqrt{{\frac{1}{21}}}} \cdot \ubrii\,
    [[\HTQ_{2},\HSQ_{1}]_{1},[\HSQ_{2},\HSQ_{1}]_{3}]_{2} \ubr+ 
   {\frac{904}{189}}\ubrii\,{\sqrt{{\frac{5}{3}}}}\ubrii\,
    [[\HTQ_{2},\HSQ_{1}]_{1},[\HSQ_{2},\HSQ_{1}]_{2}]_{2} \ubr+ 
   {\frac{856}{63}}\ubrii\,[[\HTQ_{2},\HSQ_{1}]_{1},[\HSQ_{2},\HSQ_{1}]_{1}]_{2} 
   \br- i\,{\frac{1417}{81}}\ubrii\,{\sqrt{{\frac{1}{210}}}}\ubrii\, 
    [[[\HTQ_{2},\HSQ_{1}]_{2},\HSQ_{1}]_{3},\HSQ_{1}]_{2} \ubr- 
   i\,{\frac{365}{189}}\ubrii\,{\sqrt{{\frac{1}{6}}}}\ubrii\,
    [[[\HTQ_{2},\HSQ_{1}]_{2},\HSQ_{1}]_{2},\HSQ_{1}]_{2} \br- 
   i\,{\frac{44923}{1701}}\ubrii\,{\sqrt{{\frac{1}{10}}}}\ubrii\,
    [[[\HTQ_{2},\HSQ_{1}]_{2},\HSQ_{1}]_{1},\HSQ_{1}]_{2} \ubr+ 
   i\,{\frac{2482}{1701}}\ubrii\,{\sqrt{10}}\ubrii\, 
   [[[\HTQ_{2},\HSQ_{1}]_{1},\HSQ_{1}]_{2},\HSQ_{1}]_{2} \br+ 
   i\,{\frac{166}{27}}\ubrii\,{\sqrt{{\frac{2}{15}}}}\ubrii\, 
    [[[\HTQ_{2},\HSQ_{1}]_{1},\HSQ_{1}]_{1},\HSQ_{1}]_{2} \ubr+ 
   i\,{\frac{245123}{21870}}\ubrii\,{\sqrt{{\frac{7}{3}}}}\ubrii\,
    \{\HJQ_{1},[[\HTQ_{2},\HSQ_{1}]_{1},\HTQ_{2}]_{3}\}_{2} \br- 
   {\frac{63931081}{61236}}\ubrii\,{\sqrt{{\frac{1}{105}}}}\ubrii\,
    \{\HJQ_{1},[[\HSQ_{2},\HSQ_{1}]_{3},\HSQ_{1}]_{3}\}_{2} \ubr+ 
   {\frac{914704363}{306180}}\ubrii\,{\sqrt{{\frac{1}{210}}}}\ubrii\,
    \{\HJQ_{1},[[\HSQ_{2},\HSQ_{1}]_{2},\HSQ_{1}]_{3}\}_{2} \br- 
   i\,{\frac{340927}{756}}\ubrii\,{\sqrt{{\frac{1}{21}}}}\ubrii\,
    \{\HJQ_{1},[[\HTQ_{2},\HSQ_{1}]_{2},\HTQ_{2}]_{2}\}_{2} \ubr+ 
   i\,{\frac{3401617}{6804}}\ubrii\,{\sqrt{{\frac{1}{15}}}}\ubrii\,
    \{\HJQ_{1},[[\HTQ_{2},\HSQ_{1}]_{1},\HTQ_{2}]_{2}\}_{2} \br- 
   {\frac{5162945}{20412}}\ubrii\,{\sqrt{{\frac{5}{42}}}}\ubrii\,
    \{\HJQ_{1},[[\HSQ_{2},\HSQ_{1}]_{3},\HSQ_{1}]_{2}\}_{2} \ubr+ 
   {\frac{307213}{51030}}\ubrii\,{\sqrt{{\frac{1}{6}}}}\ubrii\,
    \{\HJQ_{1},[[\HSQ_{2},\HSQ_{1}]_{2},\HSQ_{1}]_{2}\}_{2} \br- 
   {\frac{160157}{6804}}\ubrii\,{\sqrt{{\frac{1}{10}}}}\ubrii\,
    \{\HJQ_{1},[[\HSQ_{2},\HSQ_{1}]_{1},\HSQ_{1}]_{2}\}_{2} \ubr+ 
   i\,{\frac{6288067}{34020}}\ubrii\,{\sqrt{{\frac{1}{21}}}}\ubrii\,
    \{\HJQ_{1},[[\HTQ_{2},\HSQ_{1}]_{3},\HTQ_{2}]_{1}\}_{2} \br+ 
   i\,{\frac{5537}{486}}\ubrii\,{\sqrt{{\frac{1}{15}}}}\ubrii\,
    \{\HJQ_{1},[[\HTQ_{2},\HSQ_{1}]_{2},\HTQ_{2}]_{1}\}_{2} \ubr- 
   i\,{\frac{112708}{2835}}\ubrii\,\{\HJQ_{1},[[\HTQ_{2},\HSQ_{1}]_{1},\HTQ_{2}]_{1}\}_{2} 
   \br- {\frac{109177}{4860}}\ubrii\,{\sqrt{{\frac{1}{10}}}}\ubrii\, 
    \{\HJQ_{1},[[\HSQ_{2},\HSQ_{1}]_{2},\HSQ_{1}]_{1}\}_{2} \ubr- 
   {\frac{235237}{1701}}\ubrii\,{\sqrt{{\frac{1}{30}}}}\ubrii\,
    \{\HJQ_{1},[[\HSQ_{2},\HSQ_{1}]_{1},\HSQ_{1}]_{1}\}_{2} \br+ 
   i\,{\frac{617311}{81648}}\ubrii\,\{\HJQ_{1},[[\HSQ_{1},\HSQ_{1}]_{1},\HSQ_{1}]_{1}\}_{2} 
   \ubr+ i\,{\frac{262736}{1215}}\ubrii\,{\sqrt{{\frac{1}{35}}}}\ubrii\, 
    \{[\HTQ_{2},\HSQ_{1}]_{3},\HTQ_{2}\}_{2} \ubr- 
   i\,{\frac{1924}{15}}\ubrii\,{\sqrt{{\frac{2}{7}}}} \cdot \ubrii\,
    \{[\HTQ_{2},\HSQ_{1}]_{2},\HTQ_{2}\}_{2} \ubr- 
   {\frac{262924}{2835}}\ubrii\,{\sqrt{{\frac{1}{15}}}}\ubrii\,
    \{[\HTQ_{2},\HSQ_{2}]_{1},\HTQ_{2}\}_{2} \ubr+ 
   i\,{\frac{199936}{1701}}\ubrii\,{\sqrt{{\frac{2}{5}}}}\ubrii\,
    \{[\HTQ_{2},\HSQ_{1}]_{1},\HTQ_{2}\}_{2} \ubr- 
   {\frac{12316}{105}}\ubrii\,{\sqrt{{\frac{1}{5}}}} \cdot \ubrii\,
    \{[\HTQ_{2},\HSQ_{2}]_{0},\HTQ_{2}\}_{2} \ubr+ 
   i\,{\frac{839392}{945}}\ubrii\,{\sqrt{{\frac{1}{35}}}}\ubrii\,
    \{[\HSQ_{2},\HSQ_{1}]_{3},\HSQ_{2}\}_{2} \ubr- 
   i\,{\frac{5273}{63}}\ubrii\,{\sqrt{{\frac{2}{7}}}}\ubrii\,
    \{[\HSQ_{2},\HSQ_{1}]_{2},\HSQ_{2}\}_{2} \ubr+ 
   {\frac{7867}{270}} \cdot \ubrii\,{\sqrt{{\frac{1}{15}}}}\ubrii\,
    \{[\HTQ_{2},\HTQ_{2}]_{1},\HSQ_{2}\}_{2} \ubr+ 
   i\,{\frac{573379}{1134}}\ubrii\,{\sqrt{{\frac{1}{10}}}}\ubrii\,
    \{[\HSQ_{2},\HSQ_{1}]_{1},\HSQ_{2}\}_{2} \ubr+ 
   {\frac{122578}{8505}}\ubrii\,{\sqrt{{\frac{1}{3}}}}\ubrii\,
    \{[\HSQ_{1},\HSQ_{1}]_{1},\HSQ_{2}\}_{2} \ubr+ 
   {\frac{3910402}{25515}}\ubrii\,{\sqrt{{\frac{1}{35}}}}\ubrii\,
    \{[\HSQ_{2},\HSQ_{1}]_{3},\HSQ_{1}\}_{2} \ubr+ 
   {\frac{109979}{25515}}\ubrii\,\{[\HSQ_{2},\HSQ_{1}]_{2},\HSQ_{1}\}_{2} \ubr- 
   i\,{\frac{52132}{1701}}\ubrii\,{\sqrt{{\frac{2}{5}}}}\ubrii\,
    \{[\HTQ_{2},\HTQ_{2}]_{1},\HSQ_{1}\}_{2} \ubr+ 
   {\frac{4937}{2835}}\ubrii\,{\sqrt{{\frac{1}{15}}}}\ubrii\,
    \{[\HSQ_{2},\HSQ_{1}]_{1},\HSQ_{1}\}_{2} \ubr- 
   i\,3\,{\sqrt{{\frac{1}{2}}}}\ubrii\,\{[\HSQ_{1},\HSQ_{1}]_{1},\HSQ_{1}\}_{2} 
   \ubr+ i\,{\frac{243397883}{76545}}\ubrii\,{\sqrt{{\frac{1}{35}}}} \cdot \ubrii\, 
    \{(\HJQ_{1}^{2})_{2},[\HTQ_{2},\HSQ_{1}]_{3}\}_{2} \ubr- 
   i\,{\frac{993852961}{551124}}\ubrii\,{\sqrt{{\frac{1}{14}}}}\ubrii\,
    \{(\HJQ_{1}^{2})_{2},[\HTQ_{2},\HSQ_{1}]_{2}\}_{2} \ubr+ 
   {\frac{2246719}{5670}}\ubrii\,{\sqrt{{\frac{1}{15}}}} \cdot \ubrii\,
    \{(\HJQ_{1}^{2})_{2},[\HTQ_{2},\HSQ_{2}]_{1}\}_{2} \ubr- 
   i\,{\frac{61349441}{34020}}\ubrii\,{\sqrt{{\frac{1}{10}}}}\ubrii\,
    \{(\HJQ_{1}^{2})_{2},[\HTQ_{2},\HSQ_{1}]_{1}\}_{2} \ubr+ 
   {\frac{14624299}{17010}}\ubrii\,{\sqrt{{\frac{1}{5}}}} \cdot \ubrii\,
    \{(\HJQ_{1}^{2})_{2},[\HTQ_{2},\HSQ_{2}]_{0}\}_{2} \ubr- 
   i\,{\frac{420823831}{1377810}}\ubrii\,{\sqrt{{\frac{1}{2}}}}\ubrii\,
    \{(\HJQ_{1}^{2})_{0},[\HTQ_{2},\HSQ_{1}]_{2}\}_{2} \ubr- 
   {\frac{290006756}{127575}}\ubrii\,{\sqrt{{\frac{1}{35}}}} \cdot \ubrii\,
    \{\HJQ_{1},\{\HTQ_{2},\HSQ_{2}\}_{3}\}_{2} \ubr- 
   {\frac{44464001}{42525}}\ubrii\,{\sqrt{{\frac{1}{14}}}}\ubrii\,
    \{\HJQ_{1},\{\HTQ_{2},\HSQ_{2}\}_{2}\}_{2} \ubr+ 
   {\frac{4441121}{42525}}\ubrii\,{\sqrt{{\frac{1}{10}}}}\ubrii\,
    \{\HJQ_{1},\{\HTQ_{2},\HSQ_{2}\}_{1}\}_{2} \ubr- 
   i\,{\frac{42574738}{10935}}\ubrii\,{\sqrt{{\frac{1}{35}}}}\ubrii\,
    \{\HJQ_{1},\{\HTQ_{2},\HSQ_{1}\}_{3}\}_{2} \ubr- 
   i\,{\frac{160784}{945}}\ubrii\,\{\HJQ_{1},\{\HTQ_{2},\HSQ_{1}\}_{2}\}_{2} \ubr+ 
   i\,{\frac{37240771}{34020}}\ubrii\,{\sqrt{{\frac{1}{15}}}} \cdot \ubrii\,
    \{\HJQ_{1},\{\HTQ_{2},\HSQ_{1}\}_{1}\}_{2} \ubr+ 
   {\frac{61865964308}{3444525}}\ubrii\,{\sqrt{{\frac{1}{35}}}}\ubrii\,
    \{(\HJQ_{1}^{3})_{3},\HSQ_{2}\}_{2} \ubr- 
   i\,{\frac{483972673}{32805}}\ubrii\,{\sqrt{{\frac{1}{35}}}}\ubrii\,
    \{(\HJQ_{1}^{3})_{3},\HSQ_{1}\}_{2} \ubr+ 
   {\frac{19145034893}{17222625}}\ubrii\,{\sqrt{2}}\ubrii\,
    \{\{(\HJQ_{1}^{2})_{0},\HJQ_{1}\}_{1},\HSQ_{2}\}_{2} \ubr+ 
   i\,{\frac{14667913}{255150}}\ubrii\,{\sqrt{{\frac{1}{3}}}}\ubrii\,
    \{\{(\HJQ_{1}^{2})_{0},\HJQ_{1}\}_{1},\HSQ_{1}\}_{2} \ubr- 
   i\,\left({\frac{788}{45}}\ubrii\,f\, + 2\,g_1 - 
   i\,{\frac{2138885213}{9185400}}\right)\ubrii\,{\sqrt{{\frac{2}{3}}}}\ubrii\, 
    [\HTQ_{2},\HSQ_{1}]_{2} \ubr+ 
   \left( {\frac{29975713}{113400}}\,f\, 
   + {\frac{2}{3}}\,g_1 + g_2 - {\frac{253644473669}{11481750}} \right)\ubrii\, 
   {\sqrt{6}}\ubrii\,\{\HJQ_{1},\HSQ_{2}\}_{2} \ubr+ 
   i\,\left( {\frac{3106891}{136080}}\,f + 6\,g_1 + 
   {\frac{1554446111}{382725}} \right) \ubrii\,\{\HJQ_{1},\HSQ_{1}\}_{2} \, , 
$\\[5mm]
$J^P=1^-$:\\[3mm]
$
[\HBQ_0^{(3)},\HSQ_{1}]_{1} = 
8\,[[[\HTQ_{2},\HSQ_{1}]_{2},\HTQ_{2}]_{1},\HTQ_{2}]_{1} 
\ubr- 
   {\frac{598}{45}}\ubrii\,{\sqrt{{\frac{1}{35}}}}\ubrii\,
    [[[\HTQ_{2},\HSQ_{1}]_{1},\HTQ_{2}]_{3},\HTQ_{2}]_{1} \ubr+ 
   {\frac{22}{5}} \cdot \ubrii\,[[[\HTQ_{2},\HSQ_{1}]_{1},\HTQ_{2}]_{2},\HTQ_{2}]_{1} \ubr- 
   {\frac{6}{5}}\ubrii\,{\sqrt{{\frac{3}{5}}}}\ubrii\,
    [[[\HTQ_{2},\HSQ_{1}]_{1},\HTQ_{2}]_{1},\HTQ_{2}]_{1} \ubr- 
   i\,{\frac{20}{1053}}\ubrii\,{\sqrt{2}}\ubrii\,
    [[[\HSQ_{2},\HSQ_{1}]_{1},\HSQ_{1}]_{1},\HTQ_{2}]_{1} \ubr- 
   i\,{\frac{81026}{3159}}\ubrii\,{\sqrt{{\frac{2}{7}}}}\ubrii\,
    [[\HTQ_{2},\HSQ_{1}]_{3},[\HSQ_{2},\HSQ_{1}]_{2}]_{1} \ubr+ 
   i\,{\frac{480703}{15795}}\ubrii\,{\sqrt{{\frac{1}{14}}}}\ubrii\,
    [[\HTQ_{2},\HSQ_{1}]_{2},[\HSQ_{2},\HSQ_{1}]_{3}]_{1} \ubr+ 
   i\,{\frac{1156}{3159}}\ubrii\,{\sqrt{10}} \cdot \ubrii\,
    [[\HTQ_{2},\HSQ_{1}]_{2},[\HSQ_{2},\HSQ_{1}]_{2}]_{1} \ubr- 
   i\,{\frac{23098}{5265}}\ubrii\,{\sqrt{{\frac{2}{3}}}}\ubrii\,
    [[\HTQ_{2},\HSQ_{1}]_{2},[\HSQ_{2},\HSQ_{1}]_{1}]_{1} \ubr- 
   i\,{\frac{5764}{1053}}\ubrii\,{\sqrt{{\frac{2}{3}}}} \cdot \ubrii\,
    [[\HTQ_{2},\HSQ_{1}]_{1},[\HSQ_{2},\HSQ_{1}]_{2}]_{1} \ubr- 
   i\,{\frac{7126}{1755}}\ubrii\,{\sqrt{2}}\ubrii\,
    [[\HTQ_{2},\HSQ_{1}]_{1},[\HSQ_{2},\HSQ_{1}]_{1}]_{1} \ubr- 
   {\frac{1214888}{22113}}\ubrii\,{\sqrt{{\frac{1}{15}}}} \cdot \ubrii\,
    [[[\HTQ_{2},\HSQ_{1}]_{2},\HSQ_{1}]_{2},\HSQ_{1}]_{1} \ubr+ 
   {\frac{316418}{22113}}\ubrii\,{\sqrt{5}}\ubrii\,
    [[[\HTQ_{2},\HSQ_{1}]_{2},\HSQ_{1}]_{1},\HSQ_{1}]_{1} \ubr- 
   {\frac{2647}{22113}}\ubrii\,[[[\HTQ_{2},\HSQ_{1}]_{1},\HSQ_{1}]_{2},\HSQ_{1}]_{1} 
   \ubr- {\frac{96455}{7371}}\ubrii\,{\sqrt{{\frac{5}{3}}}}\ubrii\, 
    [[[\HTQ_{2},\HSQ_{1}]_{1},\HSQ_{1}]_{1},\HSQ_{1}]_{1} \ubr+ 
   {\frac{115496}{22113}}\ubrii\,{\sqrt{{\frac{1}{5}}}}\ubrii\,
    [[[\HTQ_{2},\HSQ_{1}]_{1},\HSQ_{1}]_{0},\HSQ_{1}]_{1} \br- 
   {\frac{18168323}{4914}}\ubrii\,{\sqrt{{\frac{1}{210}}}}\ubrii\,
    \{\HJQ_{1},[[\HTQ_{2},\HSQ_{1}]_{2},\HTQ_{2}]_{2}\}_{1} \ubr+ 
   {\frac{31070981}{44226}}\ubrii\,{\sqrt{{\frac{1}{6}}}}\ubrii\,
    \{\HJQ_{1},[[\HTQ_{2},\HSQ_{1}]_{1},\HTQ_{2}]_{2}\}_{1} \br+ 
   i\,{\frac{60708209}{73710}}\ubrii\,{\sqrt{{\frac{1}{21}}}}\ubrii\,
    \{\HJQ_{1},[[\HSQ_{2},\HSQ_{1}]_{3},\HSQ_{1}]_{2}\}_{1} \ubr+ 
   i\,{\frac{3473311}{22113}}\ubrii\,{\sqrt{{\frac{1}{15}}}}\ubrii\,
    \{\HJQ_{1},[[\HSQ_{2},\HSQ_{1}]_{2},\HSQ_{1}]_{2}\}_{1} \br+ 
   i\,{\frac{7691401}{1326780}}\ubrii\,
    \{\HJQ_{1},[[\HSQ_{2},\HSQ_{1}]_{1},\HSQ_{1}]_{2}\}_{1} \ubr+ 
   {\frac{10860881}{110565}}\ubrii\,{\sqrt{{\frac{1}{42}}}}\ubrii\,
    \{\HJQ_{1},[[\HTQ_{2},\HSQ_{1}]_{3},\HTQ_{2}]_{1}\}_{1} \br+ 
   {\frac{168113}{2106}}\ubrii\,{\sqrt{{\frac{5}{6}}}}\ubrii\,
    \{\HJQ_{1},[[\HTQ_{2},\HSQ_{1}]_{2},\HTQ_{2}]_{1}\}_{1} \ubr- 
   {\frac{4670303}{73710}}\ubrii\,{\sqrt{{\frac{1}{2}}}}\ubrii\,
    \{\HJQ_{1},[[\HTQ_{2},\HSQ_{1}]_{1},\HTQ_{2}]_{1}\}_{1} \br- 
   i\,{\frac{14165561}{132678}}\ubrii\,{\sqrt{{\frac{1}{5}}}}\ubrii\,
    \{\HJQ_{1},[[\HSQ_{2},\HSQ_{1}]_{2},\HSQ_{1}]_{1}\}_{1} \ubr+ 
   i\,{\frac{15262531}{265356}}\ubrii\,{\sqrt{{\frac{1}{15}}}}\ubrii\,
    \{\HJQ_{1},[[\HSQ_{2},\HSQ_{1}]_{1},\HSQ_{1}]_{1}\}_{1} \br- 
   {\frac{4125689}{398034}}\ubrii\,{\sqrt{{\frac{1}{2}}}}\ubrii\,
    \{\HJQ_{1},[[\HSQ_{1},\HSQ_{1}]_{1},\HSQ_{1}]_{1}\}_{1} \ubr- 
   {\frac{1205135}{22113}}\ubrii\,{\sqrt{{\frac{5}{6}}}}\ubrii\,
    \{\HJQ_{1},[[\HTQ_{2},\HSQ_{1}]_{2},\HTQ_{2}]_{0}\}_{1} \ubr- 
   i\,{\frac{583121}{37908}}\ubrii\,{\sqrt{{\frac{1}{5}}}} \cdot \ubrii\,
    \{\HJQ_{1},[[\HSQ_{2},\HSQ_{1}]_{1},\HSQ_{1}]_{0}\}_{1} \ubr+ 
   {\frac{3187048}{15795}}\ubrii\,{\sqrt{{\frac{1}{21}}}}\ubrii\,
    \{[\HTQ_{2},\HSQ_{1}]_{3},\HTQ_{2}\}_{1} \ubr+ 
   {\frac{7322704}{22113}}\ubrii\,{\sqrt{{\frac{1}{15}}}}\ubrii\,
    \{[\HTQ_{2},\HSQ_{1}]_{2},\HTQ_{2}\}_{1} \ubr+ 
   i\,{\frac{2149832}{36855}}\ubrii\,{\sqrt{{\frac{2}{3}}}}\ubrii\,
    \{[\HTQ_{2},\HSQ_{2}]_{1},\HTQ_{2}\}_{1} \ubr- 
   {\frac{188252}{4095}}\ubrii\,\{[\HTQ_{2},\HSQ_{1}]_{1},\HTQ_{2}\}_{1} \ubr+ 
   {\frac{1994834}{5265}}\ubrii\,{\sqrt{{\frac{1}{21}}}}\ubrii\,
    \{[\HSQ_{2},\HSQ_{1}]_{3},\HSQ_{2}\}_{1} \ubr- 
   {\frac{57338}{3159}}\ubrii\,{\sqrt{{\frac{1}{15}}}}\ubrii\,
    \{[\HSQ_{2},\HSQ_{1}]_{2},\HSQ_{2}\}_{1} \ubr+ 
   i\,{\frac{24412}{1755}}\ubrii\,{\sqrt{{\frac{2}{3}}}}\ubrii\,
    \{[\HTQ_{2},\HTQ_{2}]_{1},\HSQ_{2}\}_{1} \ubr+ 
   {\frac{169394}{5265}}\ubrii\,\{[\HSQ_{2},\HSQ_{1}]_{1},\HSQ_{2}\}_{1} \ubr- 
   i\,{\frac{365588}{3159}}\ubrii\,{\sqrt{{\frac{2}{15}}}}\ubrii\,
    \{[\HSQ_{1},\HSQ_{1}]_{1},\HSQ_{2}\}_{1} \ubr+ 
   i\,{\frac{1632412}{9477}}\ubrii\,{\sqrt{{\frac{2}{5}}}}\ubrii\,
    \{[\HSQ_{2},\HSQ_{1}]_{2},\HSQ_{1}\}_{1} \ubr- 
   {\frac{130853}{3402}}\ubrii\,{\sqrt{{\frac{1}{5}}}} \cdot \ubrii\,
    \{[\HTQ_{2},\HTQ_{2}]_{1},\HSQ_{1}\}_{1} \ubr+ 
   i\,{\frac{143203}{6318}}\ubrii\,{\sqrt{{\frac{5}{6}}}}\ubrii\,
    \{[\HSQ_{2},\HSQ_{1}]_{1},\HSQ_{1}\}_{1} \ubr+ 
   {\frac{34}{3}}\ubrii\,\{[\HSQ_{1},\HSQ_{1}]_{1},\HSQ_{1}\}_{1} \ubr+ 
   {\frac{15905009}{331695}}\ubrii\,{\sqrt{{\frac{1}{21}}}} \cdot \ubrii\,
    \{(\HJQ_{1}^{2})_{2},[\HTQ_{2},\HSQ_{1}]_{3}\}_{1} \ubr+ 
   {\frac{15220381}{66339}}\ubrii\,{\sqrt{{\frac{5}{3}}}}\ubrii\,
    \{(\HJQ_{1}^{2})_{2},[\HTQ_{2},\HSQ_{1}]_{2}\}_{1} \ubr+ 
   i\,{\frac{23890717}{331695}}\ubrii\,{\sqrt{{\frac{2}{3}}}}\ubrii\,
    \{(\HJQ_{1}^{2})_{2},[\HTQ_{2},\HSQ_{2}]_{1}\}_{1} \ubr+ 
   {\frac{67594798}{142155}}\ubrii\,\{(\HJQ_{1}^{2})_{2},[\HTQ_{2},\HSQ_{1}]_{1}\}_{1} 
   \ubr+ i\,{\frac{8311582}{66339}}\ubrii\,{\sqrt{{\frac{10}{3}}}}\ubrii\, 
    \{(\HJQ_{1}^{2})_{0},[\HTQ_{2},\HSQ_{2}]_{1}\}_{1} \ubr+ 
   {\frac{1673006}{15309}}\ubrii\,{\sqrt{{\frac{1}{5}}}} \cdot \ubrii\,
    \{(\HJQ_{1}^{2})_{0},[\HTQ_{2},\HSQ_{1}]_{1}\}_{1} \ubr- 
   i\,{\frac{101883916}{36855}}\ubrii\,{\sqrt{{\frac{1}{35}}}}\ubrii\,
    \{\HJQ_{1},\{\HTQ_{2},\HSQ_{2}\}_{2}\}_{1} \ubr- 
   i\,{\frac{5694086}{15795}}\ubrii\,{\sqrt{{\frac{1}{5}}}}\ubrii\,
    \{\HJQ_{1},\{\HTQ_{2},\HSQ_{2}\}_{1}\}_{1} \ubr+ 
   i\,{\frac{27167696}{15795}}\ubrii\,{\sqrt{{\frac{1}{5}}}}\ubrii\,
    \{\HJQ_{1},\{\HTQ_{2},\HSQ_{2}\}_{0}\}_{1} \ubr- 
   {\frac{11461781}{7371}}\ubrii\,{\sqrt{{\frac{1}{10}}}}\ubrii\,
    \{\HJQ_{1},\{\HTQ_{2},\HSQ_{1}\}_{2}\}_{1} \ubr- 
   {\frac{143216069}{66339}}\ubrii\,{\sqrt{{\frac{1}{30}}}} \cdot \ubrii\,
    \{\HJQ_{1},\{\HTQ_{2},\HSQ_{1}\}_{1}\}_{1} \ubr- 
   i\,{\frac{19275024416}{4975425}}\ubrii\,{\sqrt{{\frac{1}{21}}}}\ubrii\,
    \{(\HJQ_{1}^{3})_{3},\HSQ_{2}\}_{1} \ubr+ 
   i\,{\frac{1805647504}{1658475}}\ubrii\,{\sqrt{{\frac{1}{5}}}}\ubrii\,
    \{\{(\HJQ_{1}^{2})_{0},\HJQ_{1}\}_{1},\HSQ_{2}\}_{1} \ubr- 
   {\frac{948679282}{331695}}\ubrii\,{\sqrt{{\frac{2}{3}}}}\ubrii\,
    \{\{(\HJQ_{1}^{2})_{0},\HJQ_{1}\}_{1},\HSQ_{1}\}_{1} \ubr- 
   i\,\left({\frac{833401}{10530}}\,f\, 
   + 6\,g_1 + 
   {\frac{319403639}{51030}}\right)\ubrii\,{\sqrt{{\frac{2}{5}}}}\ubrii\,[\HTQ_{2},\HSQ_{2}]_{1} 
    \ubr- \left(
   {\frac{55751}{3510}}\,f - 2\,g_1 - 
   {\frac{109644961}{284310}}\right)\ubrii\,{\sqrt{{\frac{3}{5}}}}\ubrii\,[\HTQ_{2},\HSQ_{1}]_{1} 
   \ubr+ i\, \left( {\frac{22229104}{110565}}\,f\, 
    - 12\,g_1 - 
   {\frac{32112458159}{4975425}}\right) \cdot \br {\sqrt{{\frac{3}{5}}}}  \ubrii\, 
    \{\HJQ_{1},\HSQ_{2}\}_{1} \ubr- \left(
    {\frac{23189809}{236925}}\,f\, + 
    6\,g_1 - g_2 + {\frac{15531593255}{398034}}\right)\ubrii\, 
    {\sqrt{2}}\ubrii\,\{\HJQ_{1},\HSQ_{1}\}_{1} \, . 
$
\end{flushleft}
\end{sloppy}

\subsection*{Outlook and Conclusion}

Due to the rapid growth of the number of $\UH$-relations induced from lower
degrees and their substantial increase in complexity, carrying out the
deformation routine in the cycle of degree five will be rather
time-consuming. We anticipate that for the construction of the algebra
satisfying the postulates up to degree six very little information, possibly
none at all, has to be gained from the classical algebra in the form of
additional truly independent relations. 

The only truly independent relation of degree five, the above-mentioned
relation with spin-parity $0^+$, has been computed. It introduces 22 new
parameters in the course of the next cycle:
14 of order $\hbar$, 7 of $\hbar^2$ and one of order $\hbar^4$. In addition
to the $\UH$-relations of degree five the action of $\HBQ_0^{(3)}$ on the
truly independent relations of degree three has to be taken into account.
From our experience in the previous cycles we expect that the number of
independent parameters among the total of 25 will be reduced drastically
when going through the next cycle. 

The fact that all correspondence postulates can be consistently implemented 
up to degree five seems to be far from trivial to us. Our computations
suggest that there is an underlying structure ensuring consistency to all
degrees. To uncover this structure will be the aim of further investigations.

\end{document}